\documentstyle[preprint,prl,aps,epsf]{revtex}
\tightenlines
\begin{document}
\title{Impurity Expulsion From a Bose-Einstein Condensate}
\author{Siu A. Chin and Harald A. Forbert}
\address{
Center for Theoretical Physics, Department of Physics, Texas A\&M University, 
College Station, TX 77843}
\date{\today}
\maketitle
\begin{abstract}

When a sizable, hard-sphere-like impurity is placed in a trapped Bose-Einstein
condensate, it will create a hole in the condensate wave function. Since
it costs more energy to drill a hole at the wave function's maximum 
(near the trap center) than at its minimum (near the trap edge), the
impurity will be expelled from the condensate. This suggests that
C$_{60}$ molecules  freely falling through a condensate of atomic
hydrogen may be deflected  by a radial acceleration comparable to $g$.
The scattered fullerenes would then form a characteristic banded density
pattern outside of the trap. The detection of these deflected molecules
can be used to identify the onset of Bose-Einstein condensation.

\end{abstract}


The Bose-Einstein condensation (BEC) of particles is the macroscopic
occupation of a single quantum state. When the effect of interaction is
weak, this occupation magnifies the ground state energy by a factor $N$
equal  to the number of condensed particles. For macroscopic $N$
$\approx 10^{24}$, this extraordinarily large factor can be exploited to
magnify minute changes in the ground state. Thus to the extent
that the environment affects the condensate's ground state, BEC is
potentially an exceedingly sensitive probe of its environment.

In this work, we consider the effect of placing a hard sphere impurity
into the condensate, which forces a hole in the condensate wave
function. For translationally invariant systems, such as bulk liquid
helium, it is immaterial where this hole is located. However, for atomic
trap experiments\cite{Rb,Na,Li,H}, because the condensate wave function
is inhomogeneous, there is a position-dependent energy penalty.   This
energy penalty is ordinarily too small to be detected. We show here that
such a minute energy difference can be magnified by BEC into a
macroscopic observable phenomenon.

This effect is generic, depending only on the inhomogeneity of the
ground state wave function and its macroscopic occupation.
It is sufficient to
examine the problem in its simplest conceptual context. Consider a
non-interacting Bose gas of mass $m$ confined to a spherical cavity (an
infinite square well) of radius $b$. An impurity, regarded as a hard
sphere of radius $a$, is placed at a distance $d$ from the cavity
center. The situation is illustrated in Fig.1. The cavity traps the Bose
gas, but it may or may not also confine the impurity.  The impurity
can, but need not be microscopic. The effect is actually most pronounced
when $a$ is comparable to $b$.  Classically, there is no interaction
between the hard sphere and the cavity. However, when the cavity is
filled with a Bose gas, the ground state energy is altered by the
presence of the impurity and there is a quantum mechanically induced
interaction between the two. 

We will assume that the impurity is 
sufficiently massive that the Born-Oppenheimer approximation is
adequate. Unless noted, distances are measured in units of
$b$ and energies in units of ${\hbar^2\over{2m}}{1\over b^2}$.
The effective impurity potential is then $V(d)=N
E_0$, where $E_0$ is the ground state energy of a single particle
confined in a unit spherical cavity with an off-center hole:
\begin{eqnarray}
 -{\bf \nabla}^2  \psi({\bf r}) &&= E\psi({\bf r}),\nonumber\\
\psi({\bf r})=0\quad {\rm at}
\quad &&|{\bf r}|=1\quad 
{\rm and}
\quad |{\bf r}-{\bf d}|=a.
\label{fundeq}
\end{eqnarray}
When $d=0$, the ground state energy and
wave function are known analytically:
$E_0=\pi^2/(1-a)^2$,
$\psi_0(r)={1\over r}\sin\bigl[\pi(r-a)/(1-a)\bigr]$.
When $d>0$, spherical symmetry is broken and the problem
is non-trivial. While this can still be solved by some brute-force numerical
method, the recent suggested use of a conformal transformation\cite{conform} is
more elegant. Since cylindrical symmetry remains intact, the wave function
can be written as $\psi({\bf r})=\rho^{-1/2}{\rm e}^{im\theta}R(\rho,z)$, with
\begin{equation}
\Bigl[
-{\partial^2\over{\partial z^2}}
-{\partial^2\over{\partial \rho^2}}
+{ {m^2-{1\over 4}}\over{\rho^2} }\Bigr]R(\rho,z)=E R(\rho,z).
\label{cylindrical}
\end{equation}
Let $Z=z+i\rho$, $\tilde Z=\tilde z+i \tilde \rho$, the 
transformation\cite{complex}
\begin{equation}
\tilde Z={{Z-c}\over{1-cZ}}
\label{map}
\end{equation}
with $c$ real, will map the unit circle $z^2+\rho^2=1$ to the 
unit circle ${\tilde z}^2+{\tilde \rho}^2=1$ and the off-center 
circle $(z-d)^2+\rho^2=a^2$
to the {\it on-center} circle 
${\tilde z}^2+{\tilde \rho}^2={\tilde a}^2$ with
$\tilde a=(d+a-c)/(1-cd-ca)$, provided that
$c=(1+d^2-a^2-\sqrt{(1+d^2-a^2)^2-4d^2}\,\,)/(2d)$. Under this
transformation, both the two dimensional Laplacian and the $1/\rho^2$ term
in (\ref{cylindrical}) are transformed with the same multiplicative factor
$J(\tilde{\bf r})=[\,(1+\tilde z c)^2+(\tilde \rho c)^2\,]^2/(1-c^2)^2$,
$$
\Bigl[
-{\partial^2\over{\partial z^2}}
-{\partial^2\over{\partial \rho^2}}
+{ {m^2-{1\over 4}}\over{\rho^2} }\Bigr]
\longrightarrow 
J(\tilde{\bf r})
\Bigl[
-{\partial^2\over{\partial \tilde z^2}}
-{\partial^2\over{\partial \tilde \rho^2}}
+{ {m^2-{1\over 4}}\over{\tilde \rho^2} }\Bigr].
$$
Thus in terms of the transformed coordinates,
$\psi(\tilde {\bf r})$ satisfies
\begin{eqnarray}
J(\tilde{\bf r})(-{\tilde\nabla}^2 ) \psi(\tilde{\bf r}) 
= &&E\psi(\tilde{\bf r}),\nonumber\\
\qquad\psi(\tilde{\bf r})=0\quad {\rm at}
\quad |\tilde{\bf r}|=1\quad 
&&
{\rm and}
\quad |\tilde{\bf r}|=\tilde a.
\label{traneq}
\end{eqnarray}
This equation can be solved by expanding\cite{conform,diag}
\begin{equation}
\psi(\tilde{\bf r})=\sum_i C_i k_i^{-1}\phi_i(\tilde{\bf r}),
\end{equation}
and diagonalizing the resulting matrix
\begin{equation}
M_{ij}=k_i\langle\phi_i|J|\phi_j\rangle k_j,
\label{mat}
\end{equation}
where  $\phi_i(\tilde{\bf r})$ and $k_i^2$ are the complete set of
eigenfunctions and eigenvalues of (\ref{traneq}) with $J=1$.

The resulting ground state energies from this transformed
Hamiltonian are shown as black dots in Fig. 2. $e_0=\pi^2$ is the
ground state energy of the cavity without the impurity. The matrix (\ref{mat})
is diagonalized using 100 basis states. For all impurity sizes, the
energy is lowered as the impurity moves off-center. The intuitive reason
for this is stated in the abstract.
When the impurity approaches the
cavity surface, $d\rightarrow1-a$,  $\tilde a \rightarrow 1$, the basis
states $\phi_i$ are tightly  squeezed and the method converges slowly.
This poor convergence shows up in Fig.
2 as an unphysical up-turn in energy near $d\approx1-a$ for
$a=1/3$ and $a=1/5$. At the point of contact, the basis states are
squeezed out of existence and the method fails. Thus if the impurity is
not confined by the cavity, this method  cannot be used for $d>1-a$. For
the latter situation, the following variational trial
function gives an excellent account of the off-center energy:
\begin{equation}
\psi_{var}({\bf r})=
\bigl(1-{a\over |{\bf r}-{\bf d}|}\bigr)(1-r)
{\rm e}^{-\alpha\, \hat{\bf r}\cdot\hat{\bf d}}.
\label{var}
\end{equation}
The variational parameter $\alpha$ jumps quickly 
from zero to some $a$-dependent value for $d{>\atop\sim}0.1$.
The resulting energies for $a=1/3$ and $a=1/5$, as computed by 
the Monte Carlo method, are
shown as dash lines in Fig.2.
The energy is continuous and smooth across the contact point and is
only slightly above the diagonalization results.

When the impurity is microscopic, $a\ll 0.1$, both of the above methods
are  inefficient and the energy shift cannot be accurately computed. In
this case, we  follow the original suggestion of Huang and
Yang\cite{ppot} and replace the impurity boundary condition by a
pseudopotential. Thus instead of (\ref{fundeq}), we solve,
\begin{eqnarray}
\Bigl( -{\bf \nabla}^2 +4\pi a\delta^3({\bf r}-{\bf d})
&&{\partial\over{\partial |{\bf r}-{\bf d}|}}|{\bf r}-{\bf d}|
\Bigr)\psi({\bf r})
= E\psi({\bf r}),\nonumber\\
\qquad\psi({\bf r})=0&&\quad {\rm at}
\quad |{\bf r}|=1,
\label{pseudoeq}
\end{eqnarray}
perturbatively in powers of $a$. The result,
\begin{equation}
{E_0\over e_0}=1
+2aj^2_0(\pi d)
+(2a)^2j^3_0(\pi d)\Bigl[\cos(\pi d)-{1\over 4}j_0(\pi d)\Bigr],
\label{perturb}
\end{equation}
is shown as indicated in Fig.2.
The first order term is proportional to the cavity ground state density,
which makes precise the notion that the energy penalty is greatest at
the maximum of the ground state wave function. Note however that, for
$a\ge 1/5$, the actual energy is much steeper than the first or second
order results. The second order term above is from summing over only the
s-wave intermediate states. As pointed out by Huang and Yang\cite{ppot},
such a second order  energy shift can be positive because the
pseudopotential is not  hermitian and a singular negative term is
removed by the derivative operation on the wave function. Since we have
summed only over the s-wave intermediate states, the negative singular
term in $\psi^{(1)}_0$, which should be proportional to $1/|{\bf r}-{\bf
d}|$, started out as proportional to $1/r$. Accordingly, we removed this
contribution to the second order energy. This s-wave-only second order
energy, while in principle incomplete, is exactly correct in the limit
of $d\rightarrow 0$. As shown in Fig. 2, its inclusion yielded excellent
agreement with numerical results at all values of $d$ for $a \le 1/10$.
In comparison, the first order result is good only for $a< 1/20$.
The effect of interaction among the condensed atoms can be incorporated 
via the Gross-Pitaevskii mean-field. The result is a slight energy 
shift \cite{cmt} in accordance with the sign of the scattering length. 
For a condensate of atomic hydrogen, this effect 
is particularly negligible because of hydrogen's anomalously 
small s-wave scattering length\cite{H}.

An impurity of radius $a$ and mass $M$ will therefore be expelled from
the condensate with radial acceleration  \begin{equation}
a_r={N\over M} {\hbar^2\over{2m}}{1\over b^2}
\Bigl(-{1\over b}{{\partial E_0}\over{\partial d}}\Bigr).
\label{aradial}
\end{equation}
Since the fall-off in energy is increasingly steep as $a$ approaches 1,
the expulsion force is greater for larger impurities.
If the impurity is microscopic, with $a\ll 0.1$,
the perturbative result (\ref{perturb}) can be used. For an
order-of-magnitude estimate we can keep only the first order term and
replace the  derivative by its average value, 
\begin{equation}
a_r\approx {\pi^2\over M} {\hbar^2\over{m}}{N\over b^3}\Bigl({a\over b}\Bigr).
\label{sphere}
\end{equation}
We have restored all factors of $\hbar/m$ and $b$ so that the overall
dependence on $b$ is clear. Note that this radial acceleration is
proportional to the condensate density $\approx N/b^3$.
Currently, all experimental traps are highly
cylindrical. For these cases, we can make an equal volume
replacement: $b^3\rightarrow {3\over 4}b^2L$, where $L$ is the length of
the cylinder, 
\begin{equation}
a_r\approx {4\over 3}{\pi^2\over M}{\hbar^2\over{m}}
{N\over b^2L}\Bigl({a\over b}\Bigr).
\label{cylinder}
\end{equation}
This corresponds to an impurity expelled radially from the middle of the
cylinder. The numerical value ${4\over 3}\pi^2=13.16$ is in good agreement
with the exact cylindrical value of $4/J_1^2(z_1)=14.84$, where $z_1$
is the first zero of $J_0(z)$.

If the impurity is microscopic, the expulsion will be most
pronounced for an impurity with a large radius-to-mass ratio, 
{\it i.e.}, a hollow molecule. For  C$_{60}$, we
have $M\approx 10^{-24}$ kg and  $a\approx 4\times 10^{-10}$m.
The latest reported
Bose condensate in atomic hydrogen by Fried {\it et al} \cite{H} is ideal 
with optimal $\hbar/m=6.65\times 10^{-42}$ J-m$^2$ and 
the largest reported $N\approx10^9$. This
condensate is highly elongated with cylindrical radius $b=7.5\mu\,$m
and length $L=5$ mm.  If C$_{60}$ molecules were sprinkled over this
condensate, they would be deflected by an average radial acceleration
of 
\begin{equation}
a_r\approx 17\,{\rm m/s^2}\approx 2\, g\,.
\label{approx}
\end{equation}
For Na and Rb condensates, the effect would be reduced by a factor of
1/23 and 1/87 respectively, modulo differences in condensate density.
Since the Li condensate will collapse beyond a critical $N_c\approx
10^3$, it will not be a favored condensate to observe this effect. To
model the fullerene's additional surface attraction for hydrogen one
should properly interpret $a$ as the exact  scattering length. Thus even
if the surface attraction were to reduce the  equivalent hard sphere
radius by a factor of 2, the radial acceleration  would still be
comparable to $g$. 

The actual radial acceleration can be computed from the
cylindrical energy shift  
\begin{equation}
E_0-e_0={{\hbar^2}\over{m}}{ {4a}\over{b^2L}}
{{J_0^2(z_1 d/b)}\over{J_1^2(z_1)}}.
\label{ecyclin}
\end{equation}
Fig.3 shows the trajectories of C$_{60}$ molecules released uniformly
at a height of $6 b=45\mu$m above the cylindrical center and freely falling
through the hydrogen condensate. They are narrowly focused near
the trap's edge at $\approx 15 \mu\,m$ below the trap center. Below this
point the beam diverges but is still bounded by two caustics. The 
density distribution at $6 b=45\mu$m below the cylindrical center is shown
in Fig.4. Most of the fullerenes are expelled out of the trap forming
a sharply defined band just outside of the trap. The surprise is that
the distribution is not diffuse, but is bounded by two high density
caustics. This characteristic distribution is also generic. A similar
band is also found using harmonic oscillator wave functions. For weaker
repulsions (such as that exerted by the Na condensate), the focal plane
and the caustics would form further down from the trap, but the banded
density distribution would remain unchanged.

We emphasize that this deflection is a purely quantum mechanical
phenomenon. Classically, if C$_{60}$ molecules were raining down on such
a low  density hydrogen vapor, only one out of 100  would encounter a
single hydrogen atom. 

Since this repulsion is effective only when the atoms are condensed in
the ground state, the detection of these deflected fullerenes
can  be used to  identify the onset of Bose-Einstein condensation.
Finally, if the impurity has a negative scattering length with respect
to the condensed atoms, the impurity will be attracted and centrally
focused by
the condensate. In either case, a uniform distribution of impurities
will become patterned after passing through a condensate.

\acknowledgements
The idea of impurity expulsion was developed from considerations of impurity 
delocalization in helium droplets. The latter was first suggested by 
E. Krotscheck\cite{imp}. We have discussed\cite{cmt} impurity
expulsion in a very preliminary manner using a simple trial function and
without considering C$_{60}$ at the XXII International
Workshop on Condensed Matter Theories, Vanderbilt University, 
June 1-7, 1998. One of us (SAC) would like to thank the Institute for
Nuclear Theory at the University of Washington for hospitality 
during the Atomic Clusters
Program, July 6-24, 1998. This stay made possible the fruitful
collaboration with A. Bulgac on the use of conformal transformations 
to solve the off-center hard sphere problem. We thank E. Krotscheck, A. Bulgac, 
G. Agnolet and C.-R. Hu for helpful discussions.
This research was funded, in part, by the U. S. National Science Foundation 
grants PHY-9512428, PHY-9870054 and DMR-9509743.


\ifpreprintsty\newpage\fi
\ifpreprintsty\newpage\fi
\begin{figure}
\noindent
\vglue 0.2truein
\hbox{
\vbox{\hsize=7truein
\epsfxsize=6truein
\leftline{\epsffile{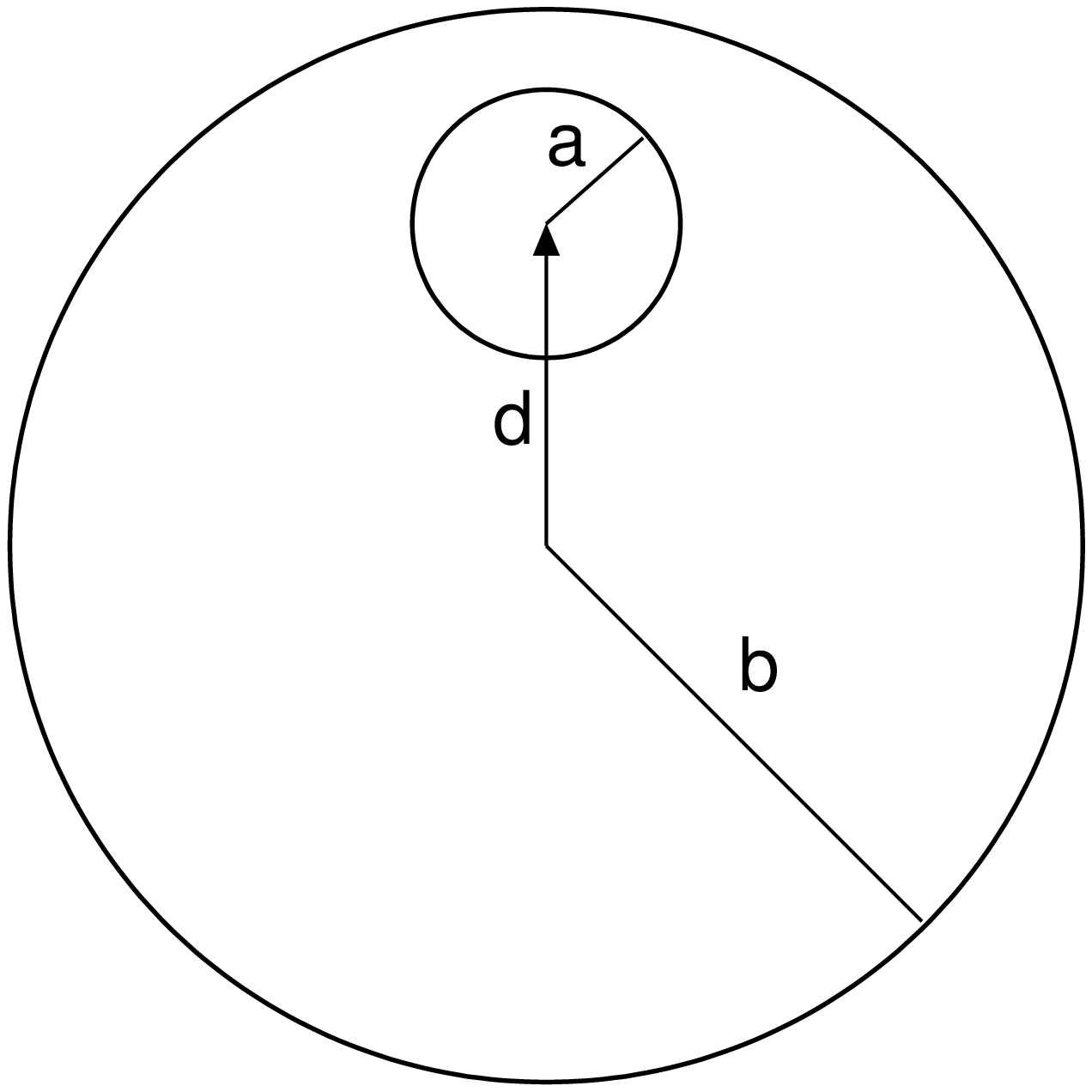}}
}
}
\caption{A Bose gas confined in a spherical cavity of
radius $b$ and excluded from an off-center impurity, which is a hard-sphere 
of radius $a$.
}
\label{fone}
\end{figure}
\ifpreprintsty\newpage\fi
\begin{figure}
\noindent
\vglue 0.2truein
\hbox{
\vbox{\hsize=7truein
\epsfxsize=6truein
\leftline{\epsffile{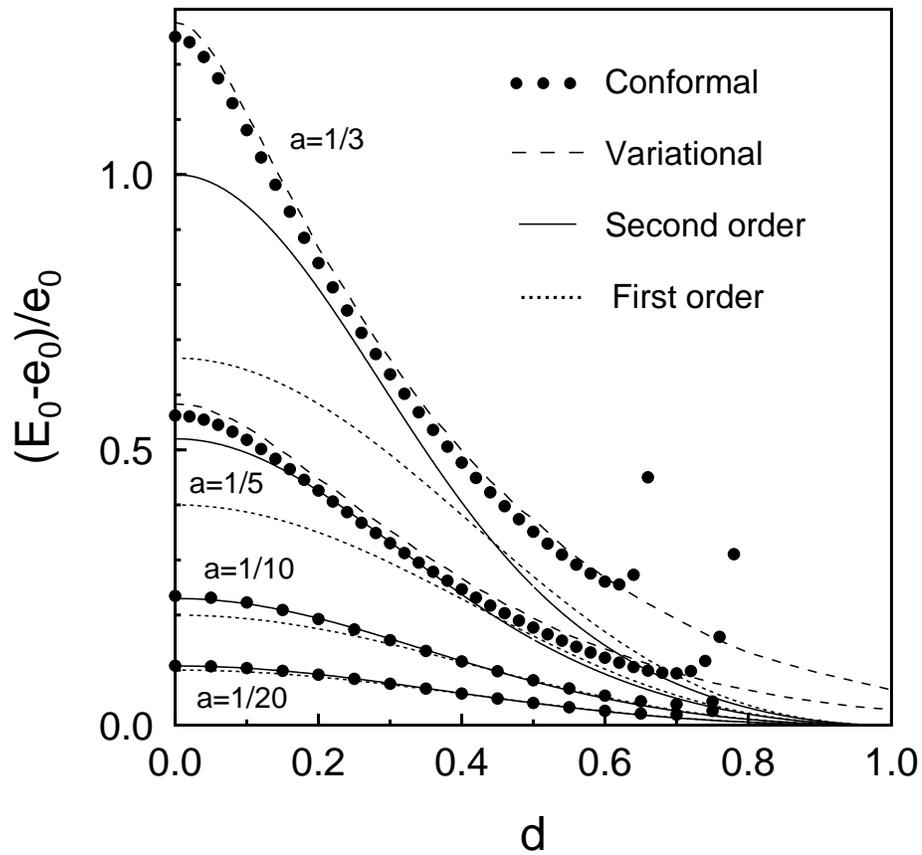}}
}
}
\caption{The ground state energy of a spherical cavity with an off-center
hard-sphere. $a$ and $d$ are measured in units of $b$. e$_0$ is the ground state
energy without the hard-sphere.
}
\label{ftwo}
\end{figure}
\ifpreprintsty\newpage\fi
\begin{figure}
\noindent
\vglue 0.2truein
\hbox{
\vbox{\hsize=7truein
\epsfxsize=6truein
\leftline{\epsffile{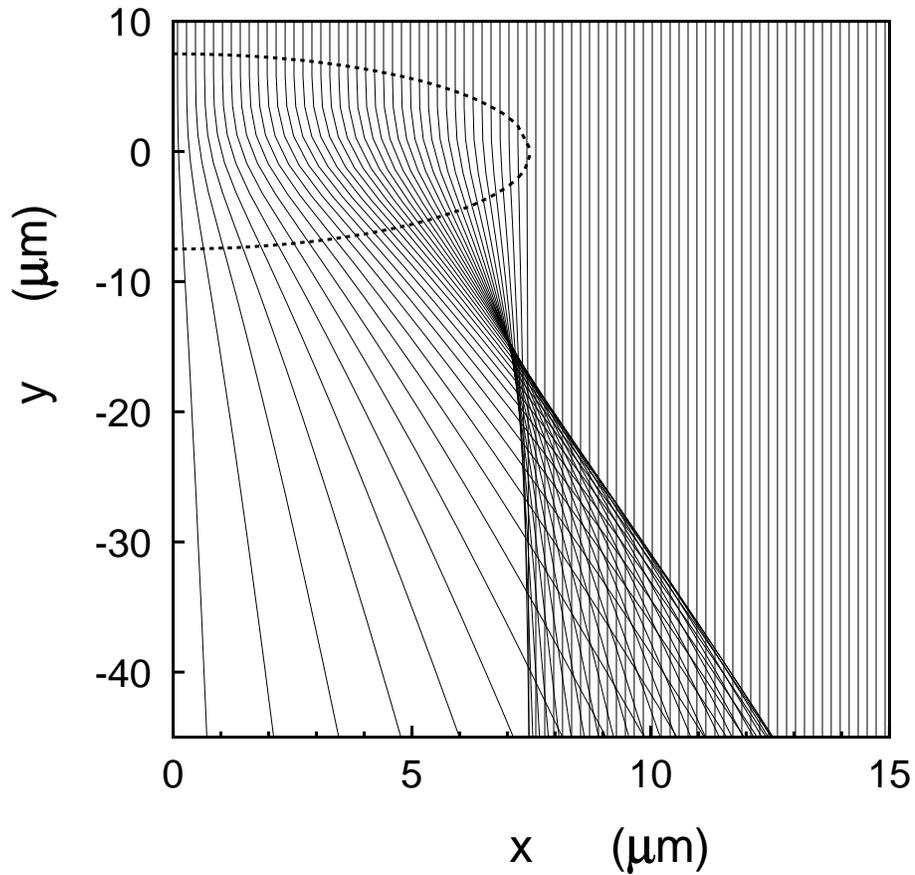}}
}
}
\caption{
Classical trajectories of C$_{60}$ molecules falling through a
cylindrical condensate of atomic hydrogen. The circular cross-section of
the cylinder is distorted by unequal x and y scales into an ellipse. See
text for details. }
\label{fthree}
\end{figure}
\ifpreprintsty\newpage\fi
\begin{figure}
\noindent
\vglue 0.2truein
\hbox{
\vbox{\hsize=7truein
\epsfxsize=6truein
\leftline{\epsffile{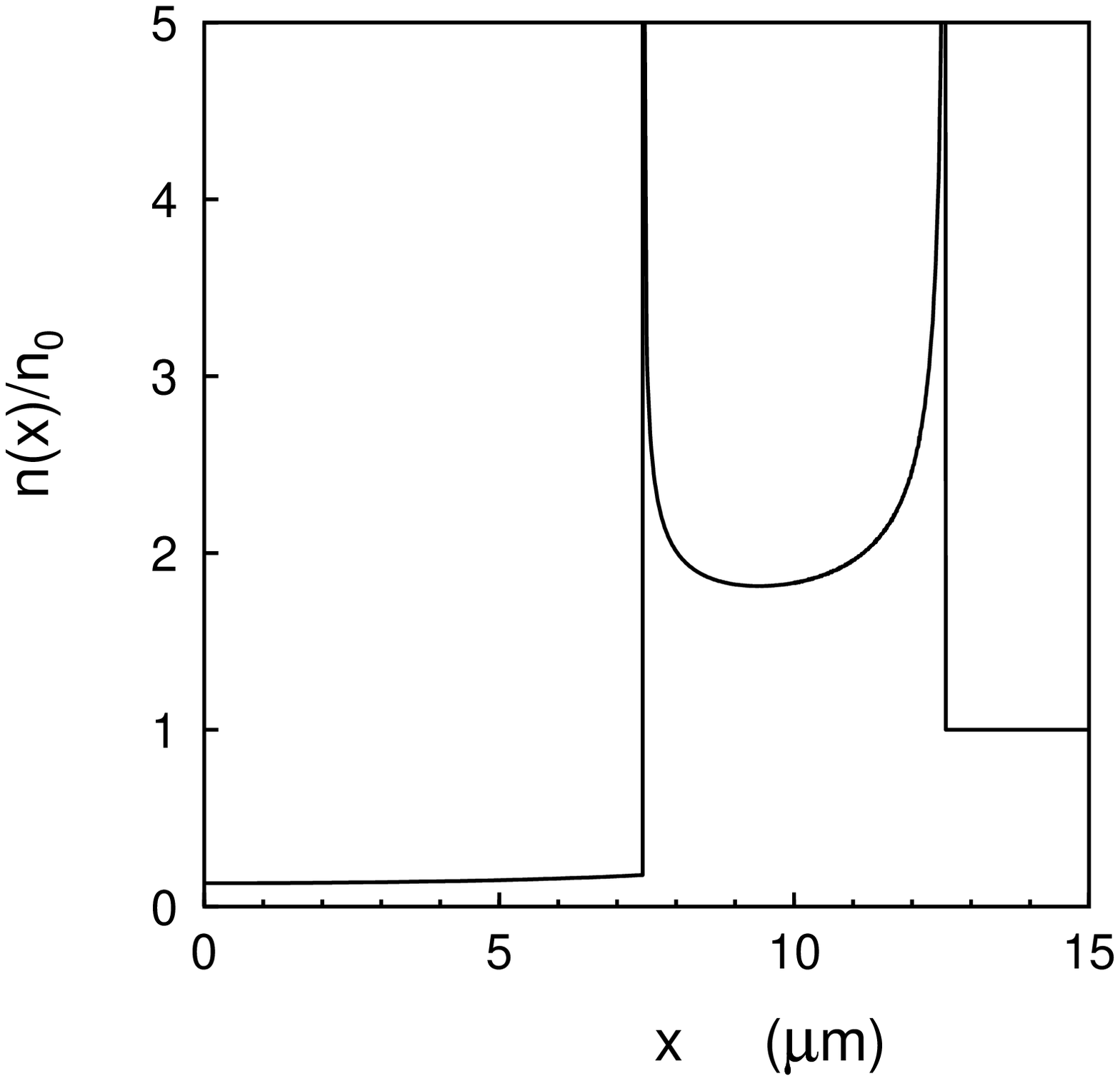}}
}
}
\caption{The density distribution of C$_{60}$ molecules falling through a cylindrical
condensate of atomic hydrogen at 45 $\mu$m below the cylindrical center.
The fullerenes are released at 45 $\mu$m above the trap center with
initial uniform density n$_0$. The condensate's cylindrical radius
is 7.5 $\mu$m.
}
\label{ffour}
\end{figure}

\begin{thebibliography}{10}
\bibitem{Rb}
M. H. Anderson, J. R. Ensher, M. R. Matthew, C. E. Wieman, and
E. A. Cornell, Science {\bf 269}, 198 (1995).

\bibitem{Na}
K. B. Davis, M.-O. Mewes, M. R. Andrews, N. J. van Druten, D. S. Durfee, 
D. M. Kurn, and W. Ketterle, Phys. Rev. Lett. {\bf 75}, 3969 (1995).

\bibitem{Li}
C. C. Bradley, C. A. Sackett, J. J. Tollet, R. G. Hulet, 
Phys. Rev. Lett. {\bf 75}, 1687 (1995).
\bibitem{H}
D. G. Fried, T. C. Killian, L. Willmann, D. Landhuis, S. C. Moss,
D. Kleppner, and T. J. Greytak, ``Bose-Einstein Condensation of Atomic Hydrogen'',
M.I.T. preprint, physics/9809017.

\bibitem{conform}
A. Bulgac, S. A. Chin, P. Magierski, Y. Yu, and H. A. Forbert, ``Fermionic
and Bosonic Bubbles and Foam", in {\it Proceedings of the International Workshop
on Collective Excitations in Fermi and Bose Systems}, Serra Negra, Brazil,
Sept.14-17, 1998, edited by C. Bertulani and M. Hussein, to be published
by World Scientific.

\bibitem{complex}
{\it Complex Analysis and Application}, P.145, by Alan Jeffrey, 
CRC Press, Boca Raton, Florida, 1992.

\bibitem{diag}
T. Prosen and M. Robnit, J. Phys. {\bf A26}, 2371 (1993).

\bibitem{ppot} 
K. Huang and C. N. Yang, Phys. Rev. {\bf 105}, 767 (1957).
            
\bibitem{cmt} S.~A. Chin and H. A. Forbert, ``Impurity Dynamics in
a Bose Condensate", to be published in ``Condensed Matter Theories'', 
Vol.14, edited by D. Ernst.

\bibitem{imp}
E. Krotscheck and S.~A. Chin,
		Chem. Phys. Lett. {\bf 227}, 143 (1994).

\end{thebibliography}
\end{document}